\documentstyle[12pt]{article}
\input epsf
\begin{document}

\newcommand{\sect}[1]{\setcounter{equation}{0}\section{#1}}
\renewcommand{\theequation}{\thesection.\arabic{equation}}

\def\rf#1{(\ref{eq:#1})}
\def\lab#1{\label{eq:#1}}
\def\nonu{\nonumber}
\def\br{\begin{eqnarray}}
\def\er{\end{eqnarray}}
\def\be{\begin{equation}}
\def\ee{\end{equation}}
\def\eq{\!\!\!\! &=& \!\!\!\! }
\def\foot#1{\footnotemark\footnotetext{#1}}
\newcommand{\nit}{\noindent}
\newcommand{\ct}[1]{\cite{#1}}
\newcommand{\bi}[1]{\bibitem{#1}}
\def\lskip{\vskip\baselineskip\vskip-\parskip\noindent}
\relax
\newcommand{\beano}{\begin{eqnarray*}}
\newcommand{\enano}{\end{eqnarray*}}
\newcommand{\hf}{\frac{1}{2}}
\newcommand{\nn}{\nonumber \\}
\font\fld=msbm10 at 12 pt
\newcommand{\fl}[1]{\mbox{\fld #1}}     
\newcommand\partder[2]{{{\partial {#1}}\over{\partial {#2}}}}
\newcommand\funcder[2]{{{\delta {#1}}\over{\delta {#2}}}}

\newcommand\me[2]{\left\langle {#1}\bv {#2} \right\rangle} 

\def\a{\alpha}
\def\b{\beta}
\def\d{\delta}
\def\D{\Delta}
\def\eps{\epsilon}
\def\vareps{\varepsilon}
\def\g{\gamma}
\def\G{\Gamma}
\def\h{{1\over 2}}
\def\l{\lambda}
\def\L{\Lambda}
\def\m{\mu}
\def\n{\nu}
\def\o{\over}
\def\om{\omega}
\def\O{\Omega}
\def\p{\phi}
\def\P{\Phi}
\def\pa{\partial}
\def\ra{\rightarrow}
\def\Ra{\Rightarrow}
\def\s{\sigma}
\def\S{\Sigma}
\def\t{\tau}
\def\th{\theta}
\def\Th{\Theta}
\def\vp{\varphi}
\def\ca{{\cal A}}
\def\cb{{\cal B}}
\def\cd{{\cal D}}
\def\ce{{\cal E}}
\def\cf{{\cal F}}
\def\cg{{\cal G}}
\def\ch{{\cal H}}
\def\cl{{\cal L}}
\def\cm{{\cal M}}
\def\cn{{\cal N}}
\newcommand\fourmat[4]{\left(\begin{array}{cc}  
{#1} & {#2} \\ {#3} & {#4} \end{array} \right)}

%
\def\PRL#1#2#3{{\sl Phys. Rev. Lett.} {\bf#1} (#2) #3}
\def\NPB#1#2#3{{\sl Nucl. Phys.} {\bf B#1} (#2) #3}
\def\NPBFS#1#2#3#4{{\sl Nucl. Phys.} {\bf B#2} [FS#1] (#3) #4}
\def\CMP#1#2#3{{\sl Commun. Math. Phys.} {\bf #1} (#2) #3}
\def\PRD#1#2#3{{\sl Phys. Rev.} {\bf D#1} (#2) #3}
\def\PRv#1#2#3{{\sl Phys. Rev.} {\bf #1} (#2) #3}
\def\PLA#1#2#3{{\sl Phys. Lett.} {\bf #1A} (#2) #3}
\def\PLB#1#2#3{{\sl Phys. Lett.} {\bf #1B} (#2) #3}
\def\JMP#1#2#3{{\sl J. Math. Phys.} {\bf #1} (#2) #3}
\def\PTP#1#2#3{{\sl Prog. Theor. Phys.} {\bf #1} (#2) #3}
\def\SPTP#1#2#3{{\sl Suppl. Prog. Theor. Phys.} {\bf #1} (#2) #3}
\def\AoP#1#2#3{{\sl Ann. of Phys.} {\bf #1} (#2) #3}
\def\PNAS#1#2#3{{\sl Proc. Natl. Acad. Sci. USA} {\bf #1} (#2) #3}
\def\RMP#1#2#3{{\sl Rev. Mod. Phys.} {\bf #1} (#2) #3}
\def\PR#1#2#3{{\sl Phys. Reports} {\bf #1} (#2) #3}
\def\AoM#1#2#3{{\sl Ann. of Math.} {\bf #1} (#2) #3}
\def\UMN#1#2#3{{\sl Usp. Mat. Nauk} {\bf #1} (#2) #3}
\def\FAP#1#2#3{{\sl Funkt. Anal. Prilozheniya} {\bf #1} (#2) #3}
\def\FAaIA#1#2#3{{\sl Functional Analysis and Its Application} {\bf #1} (#2)
#3}
\def\BAMS#1#2#3{{\sl Bull. Am. Math. Soc.} {\bf #1} (#2) #3}
\def\TAMS#1#2#3{{\sl Trans. Am. Math. Soc.} {\bf #1} (#2) #3}
\def\InvM#1#2#3{{\sl Invent. Math.} {\bf #1} (#2) #3}
\def\LMP#1#2#3{{\sl Letters in Math. Phys.} {\bf #1} (#2) #3}
\def\IJMPA#1#2#3{{\sl Int. J. Mod. Phys.} {\bf A#1} (#2) #3}
\def\AdM#1#2#3{{\sl Advances in Math.} {\bf #1} (#2) #3}
\def\RMaP#1#2#3{{\sl Reports on Math. Phys.} {\bf #1} (#2) #3}
\def\IJM#1#2#3{{\sl Ill. J. Math.} {\bf #1} (#2) #3}
\def\APP#1#2#3{{\sl Acta Phys. Polon.} {\bf #1} (#2) #3}
\def\TMP#1#2#3{{\sl Theor. Mat. Phys.} {\bf #1} (#2) #3}
\def\JPA#1#2#3{{\sl J. Physics} {\bf A#1} (#2) #3}
\def\JSM#1#2#3{{\sl J. Soviet Math.} {\bf #1} (#2) #3}
\def\MPLA#1#2#3{{\sl Mod. Phys. Lett.} {\bf A#1} (#2) #3}
\def\JETP#1#2#3{{\sl Sov. Phys. JETP} {\bf #1} (#2) #3}
\def\CAG#1#2#3{{\sl  Commun. Anal\&Geometry} {\bf #1} (#2) #3}
\def\JETPL#1#2#3{{\sl  Sov. Phys. JETP Lett.} {\bf #1} (#2) #3}
\def\PHSA#1#2#3{{\sl Physica} {\bf A#1} (#2) #3}
\def\PHSD#1#2#3{{\sl Physica} {\bf D#1} (#2) #3}
\def\PJA#1#2#3{{\sl Proc. Japan. Acad.} {\bf #1A} (#2) #3}
\def\JPSJ#1#2#3{{\sl J. Phys. Soc. Japan} {\bf #1} (#2) #3}
\def\SJPN#1#2#3{{\sl Sov. J. Part. Nucl.} {\bf #1} (#2) #3}
\begin{titlepage}
\vspace*{-1cm}
\noindent
April 1998 \hfill{US-FT/3-98}\\
\phantom{bla}
\vfill
\begin{center}
{\large\bf  Integrable Chiral Theories in $2+1$ Dimensions}
\end{center}

\begin{center}
D. Gianzo,
J.O. Madsen
and
J. S\'anchez Guill\'en
\par \vskip .1in \noindent
Departamento de F\'\i sica de Part\'\i culas,\\
Facultad de F\'\i sica\\
Universidad de Santiago\\
E-15706 Santiago de Compostela, SPAIN
\normalsize
\end{center}
\vspace{.2in}
\begin{abstract}
\vspace{.3 cm}
\small
\par \vskip .1in \noindent
Following a recent proposal for integrable theories in higher
dimensions based on zero curvature, new  Lorentz invariant submodels
of the principal 
chiral model in $2+1$ dimensions are found. They have inf\/inite local
conserved currents, which are explicitly given for the $su(2)$ case.
The construction works for any Lie algebra and in any dimension, and
it is given explicitly also for  $su(3)$. We comment on the
application to supersymmetric chiral models.

\end{abstract}
\vfill
\end{titlepage}

\sect{Introduction}
\label{sec:intro}
In this paper, we obtain Lorentz invariant submodels of the principal
chiral model in $2+1$ 
dimensions, and we find explicitly inf\/initely many conserved
currents, following a new  
generalized zero curvature approach in higher dimensions \ct{nos},
thus contributing further to its understanding.

A systematic generalization of the extremely useful zero curvature 
formulation for integrability to dimensions higher than two is a 
longstanding dif\/f\/icult problem, especially for models with Lorentz
invariance \ct{Novikov}. 
Success has been achieved so far for selfdual 
cases \ct{BZAK} and other situations which are ef\/fectively two
dimensional 
\ct{Misha}, or which are not  Lorentz invariant. 
The latter is in a strict sense the case
for the extensively studied  Modif\/ied Chiral Model \ct{Theo}, since
this model f\/ixes a direction in three dimensional spacetime
\ct{Ward}. Recently one of us with O.Alvarez and L.Ferreira \ct{nos}
proposed a geometric approach  
for $d$ dimensions based on the interplay of connections (functionals of the
f\/ields) up to rank $d-1$, where the generalized zero curvature conditions
become local f\/ield equations with the appropriate algebraic structure.
For $2+1$ dimension it was shown that for any vector $A_\m$ and
antisymmetric tensor ${B}_{\m \n}$ in a non-semisimple 
Lie algebra, the f\/latness of $A$ and constant covariance of the dual
of $B$, $D\tilde{B}=0$,  
are suf\/f\/icient conditions for generalized zero curvature and the
associated surface indepence. Once accomodated into the scheme by a
proper choice of the algebra and its representations, one can
systematically analyze 
the integrability properties of the particular models. Many of the
powerful tools of the zero curvature approach to obtain conserved
charges and solutions can thereby be used in higher dimensions.

The  $CP^1$ example was worked out this way in \ct{nos} for $A$ in
$sl(2)$ and $\tilde{B}$ in an abelian subalgebra, with  interesting
results. It was shown that the local zero curvature conditions of the
approach only {\it imply} the equations of the model for 
a special representation {\it spin} $j=1$ and that the resulting
conserved currents just correspond to the isometries of the model, 
which is therefore not integrable. On the other hand, requiring that
the equations of motion are equivalent to the zero curvature
conditions for {\em any} spin $j$, as one would like for
integrability, selects an interesting submodel \ct{ward2} with
inf\/initely many (enumerated by $j$) new conserved currents. 

In this paper we elaborate further the $CP^1$ case, obtaining general
expresions for the currents in Section \ref{sec-cp1}, and then we apply
the method to the Principal Chiral model for $su(2)$ in Section 
\ref{sec-chiral}. The formulation is even simpler in this case, and it  
leads in fact to a new class of models with inf\/initely many conserved
charges which are explicitly given. This conf\/irms further the
suggestion in \ct{nos} of the equivalence  
of zero curvature and equations of motion, implemented by
representation independence, as  
a constructive criterion of integrability, as we discuss in Section
\ref{sec-solus}. In this section we also deal with the solutions of
the submodels, working out in detail the static  
case. In section \ref{SUSY} we remark on the generalization to the
supersymmetric chiral model, and in Section \ref{sec-su3} we argue
that the result holds also for 
higher algebras and we present in detail the $su(3)$ case. In the last
Section \ref{sec-conclus}, we summarize our results and
give some general remarks on the nature of higher dimensional
integrability, as well as on applications and possible future
developments. 

\section{The ${\cal C}P^1$ model}\label{sec-cp1}

For future reference, and in order to demonstrate the method, we will 
f\/irst review the case of the ${\cal C}P^1$ model, at the same time
improving the result of \ct{nos} by giving explicit expressions for the
conserved currents (first found in \ct{japan}), and improving also the
results of \ct{japan} by finding the conserved currents in a simple
and straightforward way.  

The ${\cal C}P^1$ model is def\/ined by the equation of motion
\be
(1+|u|^2)\pa^2 u=2u^*\pa_\m u\pa^\m u\label{eq:cp1}
\ee
In order to represent the equations of motion in the form of zero
curvature conditions we follow \ct{nos}, and define the f\/ields  
\br
A_\m\eq {1\o(1+|u|^2)}
\left\{-i\pa_\m u T_+ -i\pa_\m u^*T_-+(u\pa_\m u^*-u^*\pa_\m u)T_3\right\}\nn
\tilde{B}^{(1)}_\m\eq {1\o(1+|u|^2)}
\left\{\pa_\m u P^{(1)}_{+1}-\pa_\m u^*P^{(1)}_{-1}\right\}
\er
where ${T_\pm,T_3}$ are the generators of the $sl(2)$ algebra 
\be
\left[T_+,T_-\right]=2T_3\hspace{1cm}\left[T_3,T_\pm\right]=\pm T_\pm
\ee
and $P^{(1)}_{+1}$ and $P^{(1)}_{-1}$ (together with $P^{(1)}_0$)
transform under the  
triplet representation of $sl(2)$. 
In general, we consider the commutation relations 
associated to a generic spin-$j$ representation of $sl(2)$
\br
\label{spin-j}
\left[T_3,P_m^{(j)}\right]\eq mP_m^{(j)}\nn
\left[T_\pm,P_m^{(j)}\right]\eq\sqrt{j(j+1)-m(m\pm1)}P_{m\pm1}^{(j)}\nn
\left[P_m^{(j)},P_{m^\prime}^{(j^\prime)}\right]\eq0
\er
Then, equation (\ref{eq:cp1}) is equivalent to
\be
\label{DB}
D_\m\tilde{B}^{(1)\,\m}=0
\ee
where $D_\m\cdot=\pa_\m\cdot+[A_\m,\,\cdot\,]$. The f\/ield $A_\m$ is
a f\/lat potential, 
i.e. $F_{\m\n}=0$, and so it can be put in the form $A_\m=-\pa_\m
WW^{-1}$ where $W$ is 
the group element 
\be
W=
{1\o\sqrt{1+|u|^2}}\fourmat{1}{iu}{iu^*}{1}
\ee
In order for equation (\ref{DB}) to be equivalent to the equations of
motion for the ${\cal C}P^1$ model, it is important that
$\tilde{B}^{(1)}_m$ takes its value in the {\it spin} 1
representation.  
In fact, we may define a field $\tilde{B}^{(j)}_m$ for any integer
{\it spin} $j$, by taking
\be
\tilde{B}^{(j)}_\m={1\o(1+|u|^2)}\left\{\pa_\m u P^{(j)}_{+1}-\pa_\m
u^*P^{(j)}_{-1}\right\} 
\ee
In this case, the equation $D_\m\tilde{B}^{(j)\,\m}=0$ becomes 
\br
0\eq
\left\{\sqrt{j(j+1)-2}\left(-i\pa_\m u\pa^\m
uP^{(j)}_{+2}+i\pa_\m u^*\pa^\m u^*P^{(j)}_{-2}\right)\right.\nn 
& &+\left.\left((1+|u|^2)\pa^2u-2u^*\pa_\m u\pa^\m
u\right)P^{(j)}_{+1}\right.\nn 
& &+\left.\left((1+|u|^2)\pa^2u^*-2u\pa_\m u^*\pa^\m
u^*\right)P^{(j)}_{-1}\right\} 
\er
We see that the solution of this equation is in fact independent of
$j$ for $j>1$, and that the equation is equivalent to the equation of 
motion of the ${\cal C}P^1$ model, in addition to a new condition 
(integrability condition) $\pa_\m u\pa^\m u=0$. 
Actually, this integrability condition corresponds to a known 
${\cal C}P^1$ submodel \ct{ward2}
\be
\pa^2 u=0\hspace{1cm}\pa_\m u\pa^\m u=0
\ee 

\subsection{Conserved currents}

Following again \ct{nos}, we can construct an infinite
number of conserved quantities in a simple way. They are of the form
\be
J^{(j)}_\m=W^{-1}\tilde{B}^{(j)}_\m W=\sum_{m=-j}^j J^{(j,m)}_\m
P^{(j)}_m 
\ee
In order to evaluate this expression, we use the known expression (see
for example \ct{gilmo}) for the adjoint action of a generic group
element 
\be
g=\fourmat{a}{b}{c}{d}
\ee
which is given by
\be
gP^{(j)}_ng^{-1}=\sum^j_{m=-j}\ca_{mn}P^{(j)}_m
\ee
with
\br
\ca_{mn}\eq\left[(j+m)!(j-m)!(j+n)!(j-n)!\right]^\h\nn
      & &\hspace{.8cm}\times\sum_k
	{a^{j+m-k}b^kc^{k+n-m}d^{j-n-k}\o(j+m-k)!k!(k+n-m)!(j-n-k)!}
\er
In our case, we replace $g$ in this expression by 
\be
W^{-1}={1\o\sqrt{1+|u|^2}}\fourmat{1}{-iu}{-iu^*}{1}
\ee
and we obtain expressions for the currents
\be
J^{(j,m)}_\m ={1\o(1+|u|^2)}\left\{\pa_{\m} u \ca_{m,1} -  
\pa_{\m} u^* \ca_{m,-1}\right\}\lab{jexpl}
\ee
with
\br
\ca_{m,\pm1}\eq\left[(j+m)!(j-m)!(j+1)!(j-1)!\right]^\h{1\o(1+|u|^2)^j}\nn
      & &\hspace{.8cm}\times\sum_{k=m\pm1}^{j\pm1}{(-1)^k(-i)^{\pm1-m}
	|u|^{2(k\pm1-m)}\o(j+m-k)!k!(k\pm1-m)!(j\mp1-k)!}
\er
Using these expressions, we can now calculate the currents given in
\ct{japan} in a simple way: 

\begin{itemize}
\item $m=0$
\be
J^{(j,0)}_\m=-i\sqrt{j(j+1)}{(u\pa_\m u^*-u^*\pa_\m
u)\o(1+|u|^2)^{j+1}}\sum_{k=0}^{j-1} 
\g^{(j)}_k|u|^{2k}
\ee 
\item $m>0$
\br
J^{(j,m)}_\m\eq\sqrt{(j+m)!\o
j(j+1)(j-m)!}{(-iu)^{m-1}\o(1+|u|^2)^{j+1}} \nn 
     &
&\hspace{-1cm}\times\sum_{k=0}^{j-m}\left\{\a_k^{(j,m)}|u|^{2k}\pa_\m u+ 
	(-1)^{j-m}\a_{j-m-k}^{(j,m)}|u|^{2k}u^2\pa_\m u^*\right\}
\er
\item $m<0$
\be
J_\m^{(j,m)}=(-1)^mJ_\m^{(j,-m)}
\ee
\end{itemize} 
where the coef\/f\/icients are
\br
\g^{(j)}_k\eq(-1)^k{1\o j}\left(\begin{array}{c}j \\ 
k\end{array}\right)\left(\begin{array}{c}j \\  
k+1\end{array}\right)\nn
\a_k^{(j,m)}\eq(-1)^k{n!\o(m+n-1)!}\left(\begin{array}{c}j-m \\ 
k\end{array}\right)
\left(\begin{array}{c}j+1 \\ k\end{array}\right)\nonu
\er

\section{The chiral model}\label{sec-chiral}

The pure chiral model is def\/ined by the equations
\be
\pa_\m A^\m=0\hspace{2cm} A_\m\equiv G^{-1}\pa_\m G
\ee
We take $G\in SU(2)$, but we will show in Section \ref{sec-su3} that
the construction works also in the case of higher groups. 
We observe that $A_\m$ is a f\/lat potential by construction,  
so we have the condition $F_{\m\n}=0$. We can express in a simple
way the chiral model in the general formalism of zero curvature. We
consider 
\be
\label{defTP}
A_\m=A_\m^i T_i\hspace{2cm} \tilde{B}_\m=\tilde{B}_\m^j P_j\equiv A_\m^j P_j
\ee
where $\{T_i\},\ i=1,2,3$ are the generators of the algebra $su(2)$
and we have the 
commutation relations
\br
\label{com}
\left[T_i,T_j\right]\eq i\eps_{ijk}T_k\nn
\left[T_i,P_j\right]\eq i\eps_{ijk}P_k\nn
\left[P_i,P_j\right]\eq0
\er
Using the fact that, by construction, $[A_\m,\tilde{B}^\m]=0$, it is
now easy to see that the equations of motion of the chiral model can
be represented by the zero curvature conditions 
\be
F_{\m\n}=0\hspace{2cm}D_\m\tilde{B}^\m=0
\ee
where $D_\m$ is defined after eq. (\ref{DB}).  
If we make a change of basis $\{T_i\}\longrightarrow\{T_\pm=T_1\pm
iT_2,\ T_3\}$ we 
obtain the commutation relations of $sl(2)$
\be
\left[T_+,T_-\right]=2T_3\hspace{1cm}\left[T_3,T_\pm\right]=\pm T_\pm
\ee
If we make also the change of basis 
\br
P_{\pm1}^{(1)}\eq\mp(P_1\pm iP_2)\nn
P_0^{(1)}\eq\sqrt{2}P_3
\er
then the new generators $P^{(1)}_m$ satisfy the standard commutation
relations of $sl(2)$ representations (\ref{spin-j}) for spin $j=1$.  

In this new basis
\br
A_\m\eq A_\m^-T_-+A_\m^3T_3+A_\m^+T_+\nn
\tilde{B}_\m\eq
A_\m^-P_{-1}^{(1)}+{1\o\sqrt{2}}A_\m^3P_{0}^{(1)}-A_\m^+P_{+1}^{(1)} 
\er
We may define $\tilde{B}^{(j=1)}_\m=\tilde{B}_\m$, and we wish to
generalize this to an infinite number of values of $j$, 
in such a way that the solution of
the equation $D_\m\tilde{B}^{\m\,(j)}=0$ is independent of $j$. We
find that if we define, for any integer value of $j$, the potential 
\be
\label{Bj}
\tilde{B}_\m^{(j)}=A_\m^-P_{-1}^{(j)}+{1\o\sqrt{j(j+1)}}A_\m^3P_{0}^{(j)}-
A_\m^+P_{+1}^{(j)}
\ee
Then, the condition $D_\m\tilde{B}^{\m\,(j)}=0$ reads
\br
D_\m\tilde{B}^{\m\,(j)}\eq\pa^\m A_\m^-P_{-1}^{(j)}+{1\o\sqrt{j(j+1)}}\pa^\m
A_\m^3P_{0}^{(j)}-\pa^\m A_\m^+P_{+1}^{(j)}\nn
& &\hspace{-1cm}
+\sqrt{j(j+1)-2}\left(A_\m^-A^{\m\,-}P_{-2}^{(j)}-A_\m^+A^{\m\,+}
P_{+2}^{(j)}\right)=0
\er
and we see that with this choice, the solutions of the equations are
indeed independent of $j$ for $j>1$. In fact, the zero curvature
equations  
$F_{\m\n}=0,\,D_\m\tilde{B}^{\m\,(j)}=0$ for all $j$ are equivalent to
the equations of motion for a submodel of the chiral model 
\be
\pa_\m A^\m=0\hspace{2cm}A_\m^\pm A^{\m\,\pm}=0
\ee

\subsection{Conserved currents}

Any element of $SU(2)$ can be parametrized by 
$u\in \fl{C},\,\th\in\fl{R}$ and written in the form 
\be
G={1\o(1+|u|^2)^\h}\fourmat{e^{i\th}}{u}{-u^*}{e^{-i\th}}
\label{eq:G}
\ee
With this form, the expression for the adjoint action
\be
GP^{(j)}_nG^{-1}=\sum^j_{m=-j}\ca_{mn}P^{(j)}_m
\ee
is now given by
\br
\ca_{mn}\eq\left[(j+m)!(j-m)!(j+n)!(j-n)!\right]^\hf
{e^{i\th(m+n)}\o(1+|u|^2)^j}\times\nn
      & &\hspace{1cm}\times\sum_k
	{(-1)^{k+n-m}u^mu^{*\,n}|u|^{2(k-m)}\o(j+m-k)!k!(k+n-m)!(j-n-k)!}
\er
As in the previous section we can obtain the conserved currents in the form
\be
J^{(j)}_\m=G\tilde{B}^{(j)}_\m G^{-1}=\sum_{m=-j}^j J^{(j,m)}_\m P^{(j)}_m
\ee
In the case of the chiral model this means
\be
J^{(j,m)}_\m=-\ca_{m1}A^+_\m+{1\o\sqrt{j(j+1)}}\ca_{m0}A^3_\m+\ca_{m,-1}A^-_\m
\ee
On the other hand, using (\ref{eq:G}), we have
\br
A_\m\eq{1\o(1+|u|^2)}\left\{\left[\rule{0cm}{.4cm}(u\pa_\m u^*-u^*\pa_\m
	   u)+2i\pa_\m\th\right]T_3\right.\nn
& &+\left.\left[e^{-i\th}(\pa_\m u+iu\pa_\m\th)\right]
T_++\left[-e^{i\th}(\pa_\m
	   u^*-iu^*\pa_\m\th)\right]T_-\right\}\label{eq:Achiral}
\er
Putting all together, we find explicit expressions for the conserved
currents: 

\begin{itemize}
\item $m=0$
\br
\label{J1}
J^{(j,0)}_\m\eq{(j(j+1))^{-\h}\o(1+|u|^2)^{j+1}}\sum_{k=0}^j\left\{(u\pa_\m
		u^*-u^*\pa_\m u)\left(1-{(j-k)(j+1)\o k+1}\right)\right.\nn
	& &\left.+2i\pa_\m\th\left(1+|u|^2{(j-k)(j+1)\o k+1}\right)
\right\}\l^j_k|u|^{2k}
\er
\item $m>0$
\br
\label{J2}
J^{(j,m)}_\m\eq-{e^{im\th} u^{m-1}\o(1+|u|^2)^{j+1}}\sum_{k=0}^{j-m}\left\{\left(
		{k+m\o j-k+1}+{|u|^2\o j+1}\right)\pa_\m u\right.\nn
	& &+\left({k+m\o j-k+1}-{2\o j+1}-|u|^2{j-k\o k+m+1}\right)iu\pa_\m\th\nn
	& &+\left.\left({j-k\o k+m+1}-{1\o j+1}\right)u^2\pa_\m u^*\right\}\b^{(j,m)}_k|u|^{2k}
\er
\item $m<0$
\be
\label{J3}
J_\m^{(j,m)}=(-1)^mJ_\m^{(j,-m)}
\ee
where the new coef\/f\/icients are
\br
\l^j_k\eq(-1)^k\left(\begin{array}{c} j \\ k \end{array}\right)^2\nn
\b^{(j,m)}_k\eq(-1)^k\sqrt{j+1\o j}\left[{(j+m)!\o j!}{(j-m)!\o j!}\right]^\h
\left(\begin{array}{c} j \\ k \end{array}\right)
\left(\begin{array}{c} j \\ k+m \end{array}\right)\nonu
\er
\end{itemize}

\subsection{Solutions}\label{sec-solus}

The constraints on the phase space of the theory, 
due to the inf\/inite number of conserved currents, 
gives to the theory some integrability properties that should manifest
themselves in the existence of non-trivial solutions.

Not much is known about analytic solutions or integrability properties
in dimensions greater than 2, apart from the fact that, due to
theorems like Derrick's or Coleman-Mandula \ct{DCM}, it has to be dif\/ferent
from the well studied properties of integrable theories in $d=2$. Therefore, 
the study of integrable models in $2+1$ dimensions is relevant, not
only for the models themselves, but for a deeper understanding of
integrability. 

The construction of general solutions can be attempted by the
dressing-like and other methods proposed in the generalized zero
curvature approach, which we are analyzing at present. 
In this paper, our purpose is to illustrate and check the consistency
of the method, and we will restrict ourselves to discuss the static
solutions, including the well known ones of 
the submodel of ${\cal C}P^1$. Of course, one can obtain
time dependent solutions if we make a {\em boost}, because of the
Lorentz invariant nature of the models, and one can also use the
time-independent solutions as seeds for dressing-like methods. 

We can easily verify that any solution of the Belavin-Polyakov
equations (or equivalently, the static Cauchy-Riemann equations) are
static solutions for the ${\cal C}P^1$ submodel. In particular, an
interesting solution is the baby-Skyrmion\ct{nos} 
\be
u=x+iy
\ee
More generally, many solutions for the ${\cal C}P^1$ submodel are well 
studied \ct{ward2,cp2+1}. The general solution of the equations
\be
\pa^2 u = 0\hspace{1cm}\pa^\m u\pa_\m u=0
\ee
is obtained by solving the equation
\be
x_+-tf(u)+x_-f(u)^2=g(u)
\ee
where $f$ and $g$ are two complex-analytic functions, and
$x_\pm=\h(x\pm iy)$. If we take 
$f=0$, then the equation imply that $u$ is an analytic function of
$x_+$, and thus the static solutions corresponds to the case $f=0$. 
The next simple case is $f(u)$ constant, $f(u)=k$. In this case   
\be
u=u(x_+-kt+k^2x_-)
\ee
These solutions are the {\em boosted} versions of the static 
ones\footnote{In both cases, to ensure f\/inite energy, $u$
must be a rational function}.

As we see, the variety of solutions is enormous. Some have been
investigated by numerical and approximate methods, like the so called
geodesic approximation, with complex but interesting results
\cite{ward2}. Exact time-dependent solutions, which are at present under
investigation, are very promising.   

\vspace{.5cm}

Turning now to the static solutions of the integrable chiral submodel, the
equations of motion are
\br
\pa_i A_i=0&\hspace{.5cm},\hspace{.5cm}&A_0=A_0^a T_a=0\\
A_i^\pm A_i^\pm=0&\ \ \ \Ra\ \ \ &A_i^\pm=0\hspace{1cm} i=1,2
\er
which implies 
\be
\pa_i A_i^3=0\hspace{1cm}A_i^\pm=0\label{eq:static}
\ee
Using the expression (\ref{eq:Achiral}) for $A_i^\pm$, we find
\be
\pa_i u+iu\pa_i\th=0\ \ \ \Ra\ \ \ u=ke^{-i\th}
\ee
and
\be
A_i^3={1\o(1+|k|^2)}\left\{(u\pa_i u^*-u^*\pa_i u)+2i\pa_i\th\right\}=2i\pa_i\th
\ee
so the equations of motion reduces to $u=ke^{-i\th}$ and 
\be
\nabla^2\th=0
\ee
Thus, the static solutions of the chiral submodel correspond to 
the solutions of this equation. Note that the only solutions of
this equation which are regular on the whole plane (including
infinity) are constants, which in this model implies that
there are no non-trivial static solutions with finite energy.  

It is interesting to note that the case $k=1$ (or any constant phase),
$u=e^{i\vp}$, is a solution of the ${\cal C}P^1$ model if
\be
\nabla^2\vp=0
\ee 
which is the same equation as before. This is not surprising because
there is a known relation between chiral and ${\cal C}P^1$ models. The
chiral model for $SU(2)$ is equivalent to the $O(4)$ model, and the
only static 
solutions of the $O(4)$ model are embeddings of the $O(3)$ model which
is equivalent to ${\cal C}P^1$. 

\subsection{The supersymmetric chiral model} 
\label{SUSY}

The approach used in this paper is easily generalized to the
supersymmetric chiral model. The SUSY chiral model for the group
$SU(2)$ is defined by the equations of motion
\be 
\label{susy-eom}
\cd_\a \ca^\a = 0  \quad \ca_\a = \cg^{-1}\cd_\a \cg
\ee
where $\cg$ is an $SU(2)$-valued superfield. We use the conventions of
\cite{GGRS}: the Grassman coordinates are Majorana spinors 
$\th^\a,\,\a=\pm$, which form the vector representation under the
Lorentz group $SL(2,\fl{R})$, while the space-time coordinates are
described by a symmetric, second-rank tensor 
$x^{\a\b} = (x^{++},x^{+-},x^{--})$, and 
$\cd_\a=\partder{}{\th_\a}+i\th^\b \pa_{\a\b},\,
\pa_{\a\b}x^{\s\t} = {\d_{(\a}}^\s {\d_{\b)}}^\t$. 

Defining $T_i,\,P_j$ and $\tilde{\cb}_\a$ like in equations 
(\ref{defTP}-\ref{com}), we find that the equations
\be 
\cf_{\a\b} = 0, \quad D_\a \tilde{\cb}^\a = 0,
\ee
with $D_\a\,\cdot = \cd_\a\,\cdot + [\ca_\a,\,\cdot\,]$, 
are equivalent to the equations of motion
(\ref{susy-eom}). For a spin-$j$ representation ($j\in \fl{Z}$) we
define $\tilde{\cb}_\a^{(j)}$ as in eq. (\ref{Bj}), and we find again
that the equations 
\be 
D_\a \tilde{\cb}^{\a\,(j)} = 0 
\ee
are equivalent to a submodel of the SUSY chiral model, with equations
of motion 
\be 
D_\a \ca^\a = 0 \quad\quad  \ca_\a^\pm \ca^{\a \pm} = 0 
\ee
where the $\pm$ are algebra indices, not to be confused with the spinor
indices. The expressions for the conserved currents given in equations 
(\ref{J1}-\ref{J3}) can also be applied directly in the supersymmetric
case. 

\section{The $SU(3)$ case}\label{sec-su3}

The methods described in the previous section can be generalized to
higher groups. As an example, we consider the
generalization to $SU(3)$, but it will be clear that the construction
can be generalized to any group. 

Consider f\/irst the chiral model for $SU(3)$. The equation of
motion is 
\be
\partial_\m A^\m = 0,\quad \quad A^\m = G^{-1}\partial^\m G
\ee 
with $G\in SU(3)$. The potential $A_\m$ is flat, so the corresponding
curvature is zero, $F_{\m\n}=0$. We consider 
\be 
A_\m = A_\m^i T_i, \quad \quad \tilde{B}_\m = \tilde{B}_\m^j P_j
\ee
where $T_i$ are the generators of $su(3)$, and $P_j$ tranform under
the adjoint representation of $su(3)$,
such that the total set of commutation relations are: 
\br
~[T^a,T^b]\eq {f^{ab}}_c T^c \nn 
~[T^a,P^b]\eq {f^{ab}}_c P^c \nn
~[P^a,P^b]\eq 0 
\er 
where ${f^{ab}}_c$ are the structure constants of $su(3)$. 
By construction we have $[A_\m,\tilde{B}^\m]=0$, and it is clear that
the zero curvature conditions 
\be 
D_\m \tilde{B}^\m = 0 \quad\quad F_{\m\n} = 0
\ee
are equivalent to the equations of motion. The conserved currents
which correspond to these zero curvature conditions are def\/ined by 
\be
J^\m = G \tilde{B}^\m G^{-1} = \partial^\m G G^{-1}
\ee
However, the conservation equation for $J^\m$, $\pa_\m J^\m=0$, is
completely equivalent to the equations of motion, $\pa_\m A^\m=0$, and
we do not find any new conserved currents.  

\subsection{An integrable submodel of the su(3) chiral model} 

In order to find an integrable submodel of the $su(3)$ chiral model,
we will define fields $\tilde{B}_\L^\m$ for an infinite set of
highest weight representations of $su(3)$, with highest weight $\L$,
such that the conditions 
\be 
D_\m \tilde{B}_\L^\m = 0 \quad\quad F_{\m\n}=0
\ee  
are equivalent to the equation of motion of the submodel. 
As in the previous sections, we will do this by def\/ining 
\be
\tilde{B}_\L = A_\m^i P^\L_i 
\ee
where $P^\L_i$ belongs to a certain subset of the generators of the
representation. 

As a generalization of $su(2)$ representations with integer spin, we
consider $su(3)$ representations with a highest weight of the
form $\L=n\,(\a_1+\a_2)$, where $\a_i$ are the simple roots.
As an illustration we show in figure 1 the weight diagram for the case
$n=2$, which is the $\bf \{ 27 \}$ of $su(3)$, but the construction
given below works for general $n$. 

\begin{figure}
\begin{center}
\hbox{\hspace{30mm}\epsfxsize=280pt \epsfbox{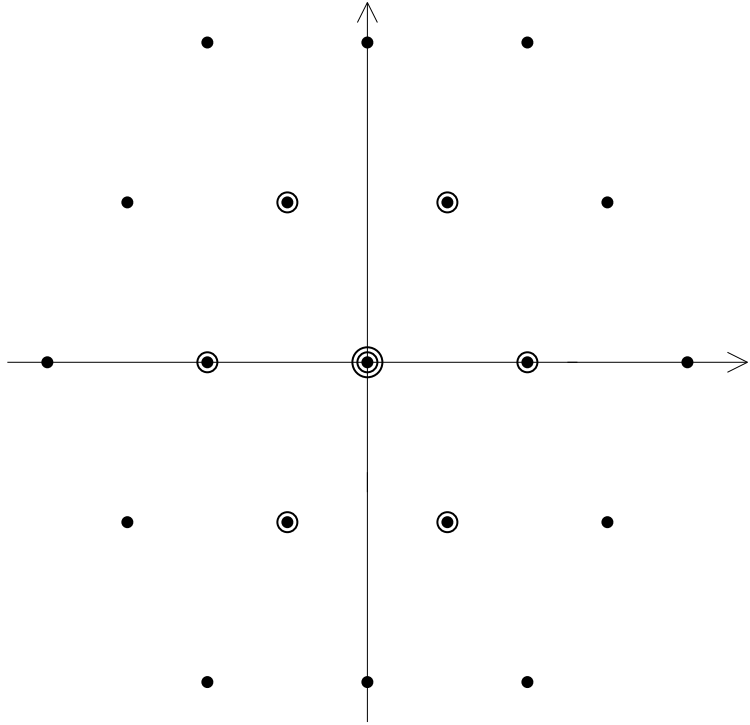}} 
\end{center}
\caption{weight diagram for the $\bf \{ 27 \}$ of $su(3)$. 
The rings indicate multiplicities.}
\end{figure}

Before we proceed to the definition of the field $\tilde{B}_\L^\m$, we
must define the subset $P^\L_i$ of the generators, and we will show
that the chosen subset have certain nice commutation relations with
the generators of the $su(3)$ algebra. 

It is convenient to make a transformation to a basis with generators
$H_i=\a_i\cdot H$, $E_\a$, and commutation-relations:  
\br
{}[H_i,E_\b] \eq \a_i\cdot \b\, E_\b \nn
{}[E_\a,E_{-\a}] \eq \a\cdot H \nn
{}[E_\a,E_\b] \eq N_{\a \b}
E_{\a+\b}\,\,\,\mbox{ for } (\a+\b)\in \Delta_R 
\er 
where $\Delta_R$ is the set of roots. 
We choose to take $E^\dagger_{\a} = E_{-\a}$. 
We take a basis for the representation of the form: 
$P_{(\l,j,m)}$, where $\lambda$ is a weight and $(j,m)$ are the $su(2)$
quantum numbers corresponding to a decomposition of the representation
in terms of representations of the $su(2)$-subalgebra generated by 
$\{(\a_1\!+\!\a_2)\cdot H, E_{\pm (\a_1+\a_2)}\}$. 
Def\/ine: 
\br
\label{defP}
P_1 = \a_1 \cdot P \eq \frac{1}{2} (x \,P_{(0,0,0)} + y\, P_{(0,1,0)}) \nn
P_2 = \a_2 \cdot P \eq \frac{1}{2} (x \,P_{(0,0,0)} - y\, P_{(0,1,0)}) \nn
P_{\pm(\a_1+\a_2)} \eq \mp \frac{1}{2} 
x\, [E_{\pm(\a_1+\a_2)}, P_{(0,1,0)}] \nn
P_{\pm\a_i} \eq \mp \frac{1}{2} [E_{\pm \a_i},P_i] 
\er
where $x$ and $y$ are constants to be defined later. 
It is easy to show the following relations: 
\br 
\label{com1}
{}[E_\a,P_{-\a}] + [E_{-\a},P_{\a}] \eq 0 \nn
{}[E_{\pm \a_i},P_i] + [H_i,P_{\pm \a_i}] \eq 0
\nn
~[E_{\pm (\a_1+\a_2)},(\a_1+\a_2)\cdot P]
+ [(\a_1\!+\!\a_2)\cdot H,P_{\pm (\a_1\!+\!\a_2)}] \eq 0 \nn
{}[E_{\pm (\a_1+\a_2)},(\a_1\!-\!\a_2)\cdot P]
+ [(\a_1\!-\!\a_2)\cdot H,P_{\pm (\a_1+\a_2)}]  =  0
\er 
where the last relation is a consequence of the fact that 
$(\a_1\!-\!\a_2)\cdot P$ is an $su(2)$-singlet. 

Using some algebra, it is possible to show that 
\be
[E_{\a_1+\a_2},[E_{\a_1+\a_2},P_{-\a_1}]+
[E_{-\a_1},P_{\a_1+\a_2}]] = 0,
\ee
which implies  
\be
\label{xy}
[E_{\a_1+\a_2},P_{-\a_1}]+[E_{-\a_1},P_{\a_1+\a_2}] 
\propto P_{(\a_2,\frac{1}{2},\frac{1}{2})}. 
\ee
Inserting the definitions (\ref{defP}) into this equation, we find
that the left hand side is a sum of two terms proportional to
respectively $x$ and $y$, and we can therefore choose these constants
in such a way that 
\be
[E_{\a_1+\a_2},P_{-\a_1}]+[E_{-\a_1},P_{\a_1+\a_2}] 
=0
\ee
Furthermore, we f\/ind that 
\br
0\eq[E_{-(\a_1+\a_2)},[E_{\a_1+\a_2},P_{-\a_1}]+[E_{-\a_1},P_{\a_1+\a_2}] ] \nn
 \eq[E_{-(\a_1+\a_2)},P_{\a_2}]+[E_{\a_2},P_{-(\a_1+\a_2)}] 
\er
These two relations, together with $E^\dagger_\a = E_{-\a}$,
show that 
\be
\label{com2}
[E_{\pm(\a_1+\a_2)},P_{\mp\a_i}]+
[E_{\mp\a_i},P_{\pm(\a_1+\a_2)}] = 0
\ee

The generators $P_\a$ and $P_i$ are the subset of generators
that we where looking for. Indeed, writing $A_\m$ in the form
$A_\m = (A^\a)_\m E_\a + (A^i)_\m H_i$, we
def\/ine: 
\be
(\tilde{B}_\L)_\m = (A^\a)_\m P_\a + (A^i)_\m P_i
\ee
Using the commutation relations (\ref{com1}) and (\ref{com2}) we find
that the equations
\be 
D_\m \tilde{B^\m} = 0, \quad\quad F_{\m\n} = 0 
\ee
are equivalent to the equations of motion of the chiral model, in
addition to the constraints
\br 
\label{con}
(A^1)_\m (A^{\pm \a_2})^\m \eq 0 \nn
(A^2)_\m (A^{\pm \a_1})^\m \eq 0 \nn
(A^{\a_1})_\m (A^{\a_2})^\m \eq 0 \nn
(A^\a)_\m (A^\b)^\m \eq 0 \,\,\,\mbox{ for }(\a+\b)\in
\D_w\setminus\D_R
\er
where $\D_w$ is the set of weights. 
We see that the constraints are independent of the choice of highest
weight $\L$, and so this set of constraints defines an integrable
submodel of the chiral $SU(3)$ model, in which the zero curvature
conditions give rise to an infinite number of conserved currents. 

\section{Conclusions}\label{sec-conclus}

Our results confirm the validity, and illustrate the simplicity, 
of the new higher dimensional zero
curvature approach to investigate integrability and construct Lorentz
invariant models with infinitely many conserved charges. This is the
main conclusion here, in  
addition to the particular interest of the cases considered; the
Principal chiral model and the $O(3)$ model has both been extensively
studied. 

In order to compare our results with the results available in the
literature, we have in this paper focused on 2+1
dimensions. Notice, however, that the dimensionality of space-time is
never used in our construction, and in fact all the results in this
paper are immediately applicable in any number of dimensions. 

Integrability in two dimensions is quite well understood at
present, but it has many peculiar features which cannot be transferred to
higher dimensions. It is therefore of great importance to improve our 
understanding of the nature of integrability beyond two dimensions. 
For this, it will be very useful to have a better understanding of  
the solutions of these models, including time-dependent solutions 
which can not be obtained by boosting static solutions. 
A general analysis of these solutions, as well as a study of several
interesting theories in 3+1 dimensions, is in progress. 

\vspace{1 cm}

\section*{Acknowledgements} 

We would like to thank J. Luis Miramontes for interesting
discussions. We are supported by EC contract 
TMRERBFMRXCT960012 and by DGICYT PB 96-0960.

\end{document}